\documentclass[conference]{IEEEtran}

\IEEEoverridecommandlockouts
\usepackage{cite}
\usepackage{amsmath,amssymb,amsfonts}
\usepackage{algorithmic}
\usepackage{graphicx}
\usepackage{textcomp}
\usepackage{xcolor} 
\usepackage[pages=some,placement=top]{background}
\def\BibTeX{{\rm B\kern-.05em{\sc i\kern-.025em b}\kern-.08em
    T\kern-.1667em\lower.7ex\hbox{E}\kern-.125emX}}
\begin{document}


\title{Cellular Traffic Prediction with \\Recurrent Neural Network}

\author{\IEEEauthorblockN{Shan Jaffry}
\IEEEauthorblockA{\textit{DGUT-CNAM Institute } \\
\textit{Dongguan University of Technology, Dongguan, China}\\
s.jaffry@dgut.edu.cn}
}

\maketitle

\begin{abstract}
Autonomous prediction of traffic demand will be a key function in future cellular networks. In the past, researchers have used statistical methods such as Autoregressive integrated moving average (ARIMA) to provide traffic predictions. However, ARIMA based predictions fail to give an exact and accurate forecast for dynamic input quantities such as cellular traffic. More recently, researchers have started to explore deep learning techniques, such as, recurrent neural networks (RNN) and long-short-term-memory (LSTM) to autonomously predict future cellular traffic. In this research, we have designed a LSTM based cellular traffic prediction model. We have compared the LSTM based prediction with the base line ARIMA model and vanilla feed-forward neural network (FFNN). The results show that LSTM and FFNN accurately predicted the future cellular traffic. However, it was found that LSTM train the prediction model in much shorter time as compared to FFNN. Hence, we conclude that LSTM models can be effectively even used with small amount of training data which will allow to timely predict the future cellular traffic.
\end{abstract}


\section{Introduction}

Cellular communication is the most popular and ubiquitous telecommunication technology. Recently, owing to novel use cases, such as, multimedia video download, 4K/8K streaming etc., the amount of cellular data traffic has soared exponentially. It is expected that in the near future, i.e. by 2023, the monthly mobile data demands will exceed beyond 109 Exabyte (Exa = $10^{18}$)  which currently rests at a modest 20 Exabytes per month consumption \cite{cerwall2018ericsson}. Cellular users, nevertheless, will expect high speed and ubiquitous connectivity from the network operators. 
Providing unhindered, ubiquitous, and high quality of service will be a serious challenge for network operators. Network operators must update traffic planning tools so they can know in advance about the state of future traffic demands. Hence, operators will rely on data-driven self-organizing networks (SON) powered by machine learning (ML) and artificial intelligence (AI). ML and AI enabled networks can preemptively take important decisions with limited human intervention. Prediction of cellular and data traffic patterns will be a key job that SON perform. 

Cellular traffic prediction will enable network operators to promptly distribute resources as per the requirement of competing users. 
With the informed network state, operators may also allow resource sharing between devices \cite{jaffry2018effective,jaffry2018shared}. This will also enable high spectral efficiency and will prevent outages caused due to cell overload. If a network can accurately predict future traffic loads in specific cells, it may take preventive actions to avoid outages. For example, network may permit device-to-device communication to relieve the base station  \cite{jaffry2017distributed}.

\begin{figure}[t]
	\centering
	\includegraphics[width=3.2 in, trim={6.0cm 2.3cm 6.5cm 2.0cm},clip = true]{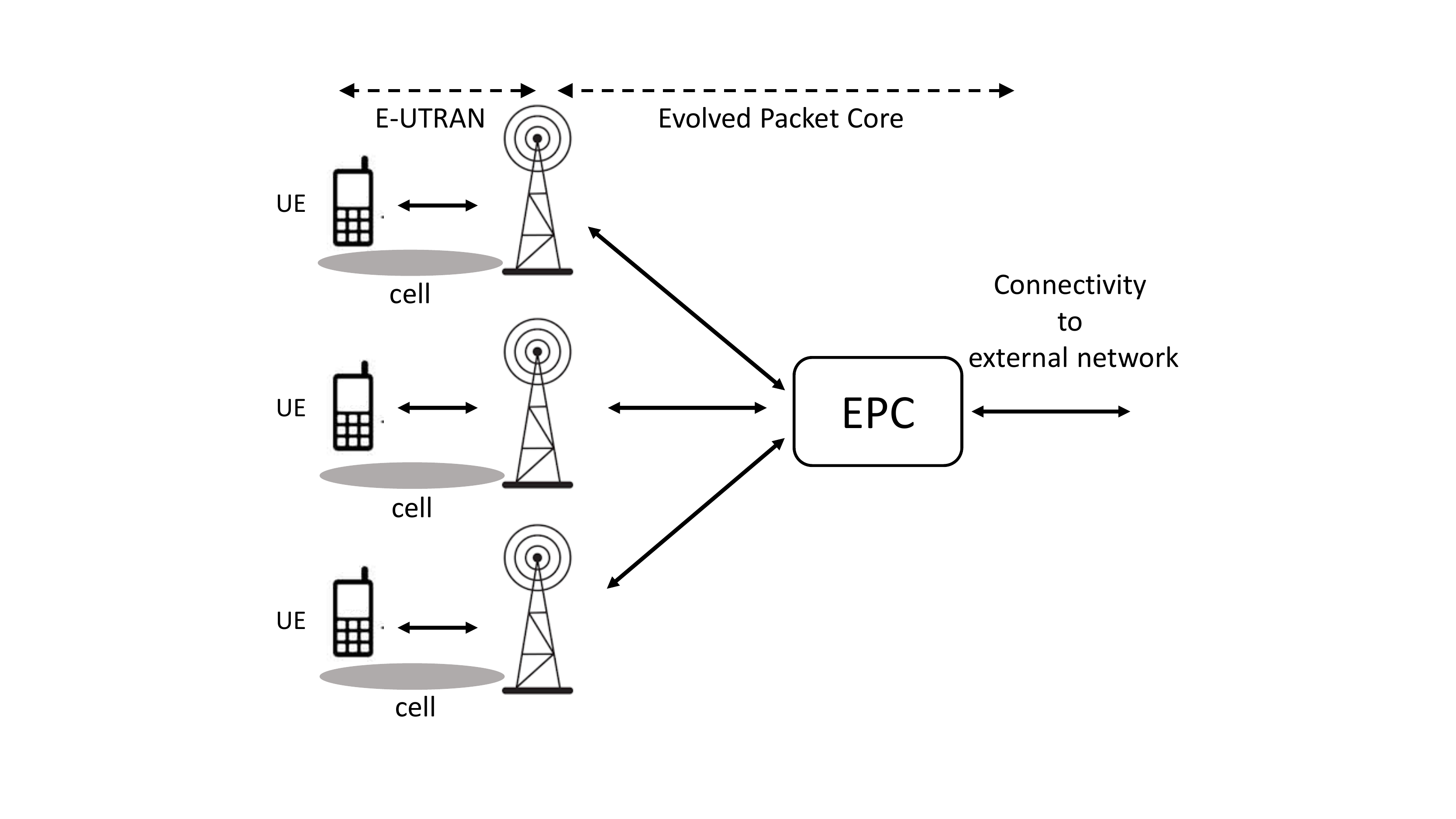}	
	\caption{LTE-A network architecture.}
	\label{fig:LTEArchitecture}
\end{figure}

Recent advances in data analytics and the availability of powerful computing machines have enabled operators to harness the power of Big data to analyze and predict network operations. Hence, advanced variations of data driven ML and AI techniques are playing an ever increasing role in all aspects of modern human lives. In particular, deep learning, a special class of ML and AI algorithms, can solve enormously complex problems by leveraging the power of very deep neural network layers \cite{lecun2015deep}.  
Deep learning algorithms can extract valuable feature information from the raw data to predict outcomes. Deep learning has made great strides recently due to advent of user friendly libraries and programming environment such as Tensorflow, Keras, PyTorch, Pandas, and Scikit etc. Deep learning algorithms such as recurrent neural network (RNN), convolutional neural network (CNN) etc. are being extensively used in application, such as, computer vision \cite{voulodimos2018deep}, health informatics \cite{ravi2016deep}, speech recognition \cite{deng2013new}, and natural language processing \cite{young2018recent} etc. In future, it is anticipated that majority operations in sixth generation (6G) cellular network will be solely catered by AI and deep learning algorithms \cite{zappone2019wireless}.  

An AI-enabled SON network will perform long and short term analysis on the data obtained from the end users and/or network \cite{chen2019artificial}. This self-optimization will reduce the over all capital expenditures (CAPEX) and operational expenditure (OPEX) required for network planning and maintenance. For example, a key issue concerning increasing CAPEX and OPEX for service providers is the identification and remedy of anomalies that may arise within a cell.  
To learn and prevent the cell from going into the anomalous state, it is necessary for the network to predict the future traffic demands. 

\begin{figure}[t]
	\centering
	\includegraphics[width=4.2 in, trim={10.2cm 5.0cm 9cm 5.2cm},clip = true]{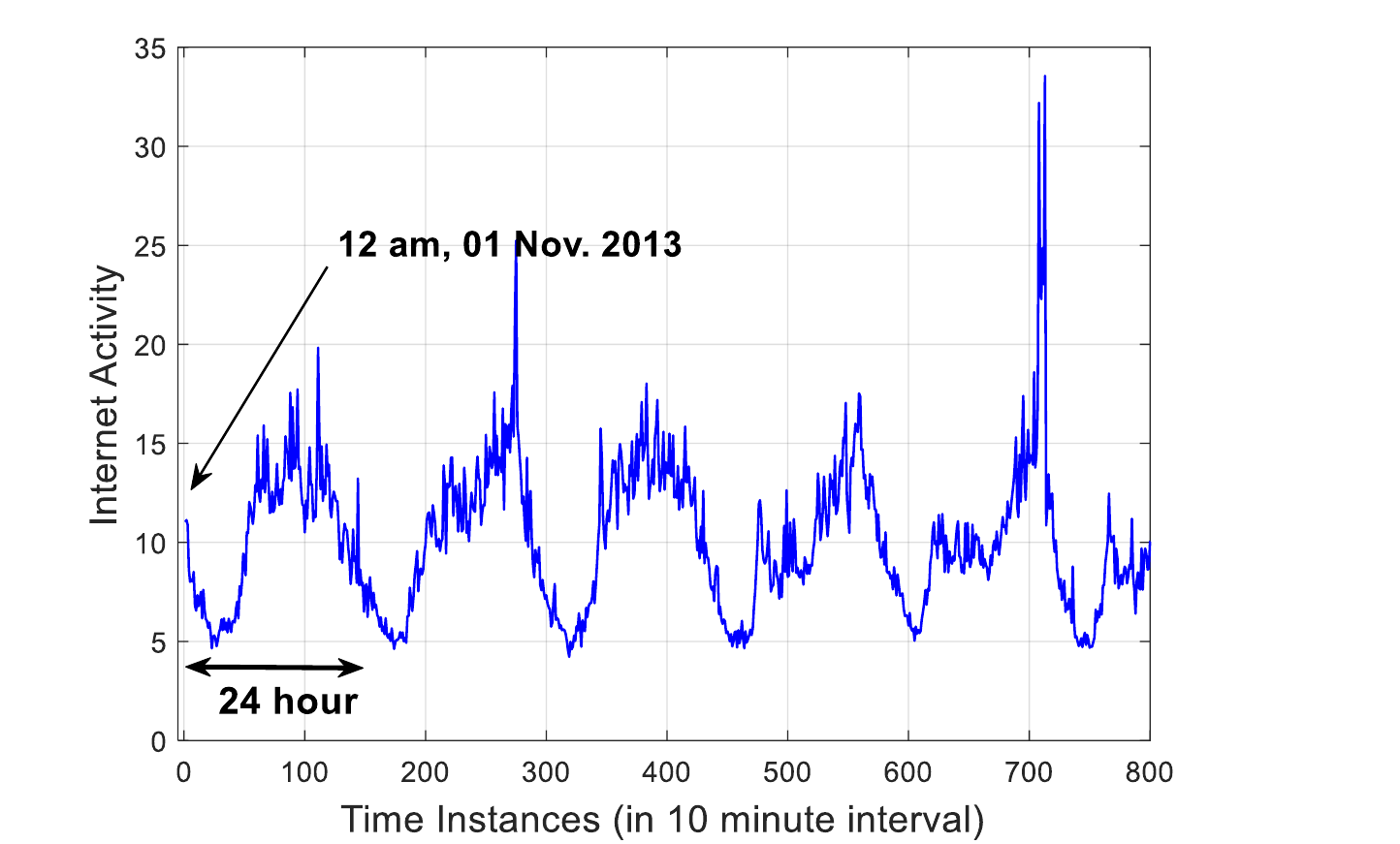}	
	\caption{Grid 01 Internet Activity for first 5 days.}
	\label{fig:Activty}
\end{figure}

In the past researcher have proposed to forecast the cellular traffic using statistical models, such, as Autoregressive integrated moving average (ARIMA) and its variants \cite{shu2005wireless}. A known limitation of ARIMA is that it reproduce the time series patterns based on average of the past values. However, ARIMA may fail to accurately predict traffic patterns in highly dynamic environments such as cellular network. Nevertheless, ARIMA can give a descent estimate of future traffic and may serve as a baseline prediction model.

Recently, deep learning based techniques to forecast any time series traffic is getting more popular. For cellular applications, deep learning techniques learn the past history of network traffic to train models such as vanilla feed-forward neural network (FFNN), recurrent neural network (RNN), or long-short-term-memory (LSTM) etc. In \cite{qiu2018spatio}, researchers have proposed to use RNN with multi-task learning to design a spatio-temporal prediction model for cellular networks. Researchers in \cite{zhao2019celltrademap} have applied neural network models on cellular traffic to analyze trade activities in urban business districts. A comparative study between LSTM and ARIMA models was conducted by researchers in \cite{azari2019cellular}. 

Inspired by the works presented earlier, in this paper we will use the real world call data record to forecast future cellular traffic using LSTM. 
In particular, we will compare our results with the ARIMA model and vanilla feed forward neural network (FFNN) models. We will demonstrate that LSTM models learn the traffic patterns very quickly as compared to FFNN and ARIMA models.

The rest of the paper is organized as follows. The system model is presented in Section \ref{system_model}. The cellular traffic prediction model is presented in Section \ref{sec_traffic_prediction}. We discuss the results in Section \ref{sec_results} followed by conclusion in Section \ref{sec_conclusion}.

\section{System Model}
\label{system_model}

Figure \ref{fig:LTEArchitecture} shows the our system model which comprise of Long Term Evolution - Advanced (LTE-A) network. 
The architecture of LTE-A is broadly categorized into three layers. The core network (CN), the access network, and the end user equipment (UE) \cite{elnashar2014design}. 

The wireless communication take place between a UE and evolved NodeB (eNB) over the access network which is called UMTS terrestrial radio access network (E-UTRAN) in LTE-A nomenclature. The core network, which is formally known as evolved packet core (EPC), makes essential the network level decisions. The EPC further contain several logical entities such as serving gateway (SGW), packet data network gateway (PGW), and mobility management unit (MMU) etc. Detailed explanation of these logical entities and LTE-A architecture is out of scope of current paper. Readers can refer relevant materials, for example \cite{elnashar2014design}. The call data record (CDR) that we will use in this research was gathered at the EPC level layer. The execution of LSTM predictive model will also take place at this layer. 

\subsection{Data Record Details}

The call data record used in this research was published by Telecom Italia for Big Data Challenge competition  \cite{telecom_ItaliaCDR}. 
Telecom Italia collected cellular and internet activities of its subscribers within the city of Milan in Italy. 
In the CDR, Milan city is divided into $100 \times 100$ square grids. 
Each grid has a length of 0.235 Km and an area of 0.055 Km$^2$. 
The data record has been collected for 62 days, starting from 1st November 2013 till  1st January 2014. 
Data for the single day is stored in a single file which means that there are 62 files in the dataset. 
Readers can refer to \cite{parwez2017big} for detailed explanation on the CDR. 

The spatio-temporal CDR contains following  fields.

\begin{itemize}
	\item  Grid ID. 
	\item  Time Stamp: Raw timestamp was recorded in milliseconds units with the interval of 10 minutes.
	\item  Country code.
	\item  Inbound SMS Activity: Indicates the incoming SMS activity in a particular grid observed within 10 minute interval.
	\item  Outbound SMS Activity: Indicates the outgoing SMS activity in a particular grid observed within 10 minute interval.
	\item  Inbound Call Activity: Indicates the incoming calling in a particular grid observed within 10 minute interval.
	\item  Outbound Call Activity: Indicates the outgoing calling activity in a particular grid observed within 10 minute interval.
	\item  Internet Activity:  Indicates the internet usage by cellular users in a particular grid observed within 10 minute interval.\\
\end{itemize}

\begin{figure}[t]
	\centering
	\includegraphics[width=3.0 in, trim={6.5cm 2.5cm 5.5cm 2.0cm},clip = true]{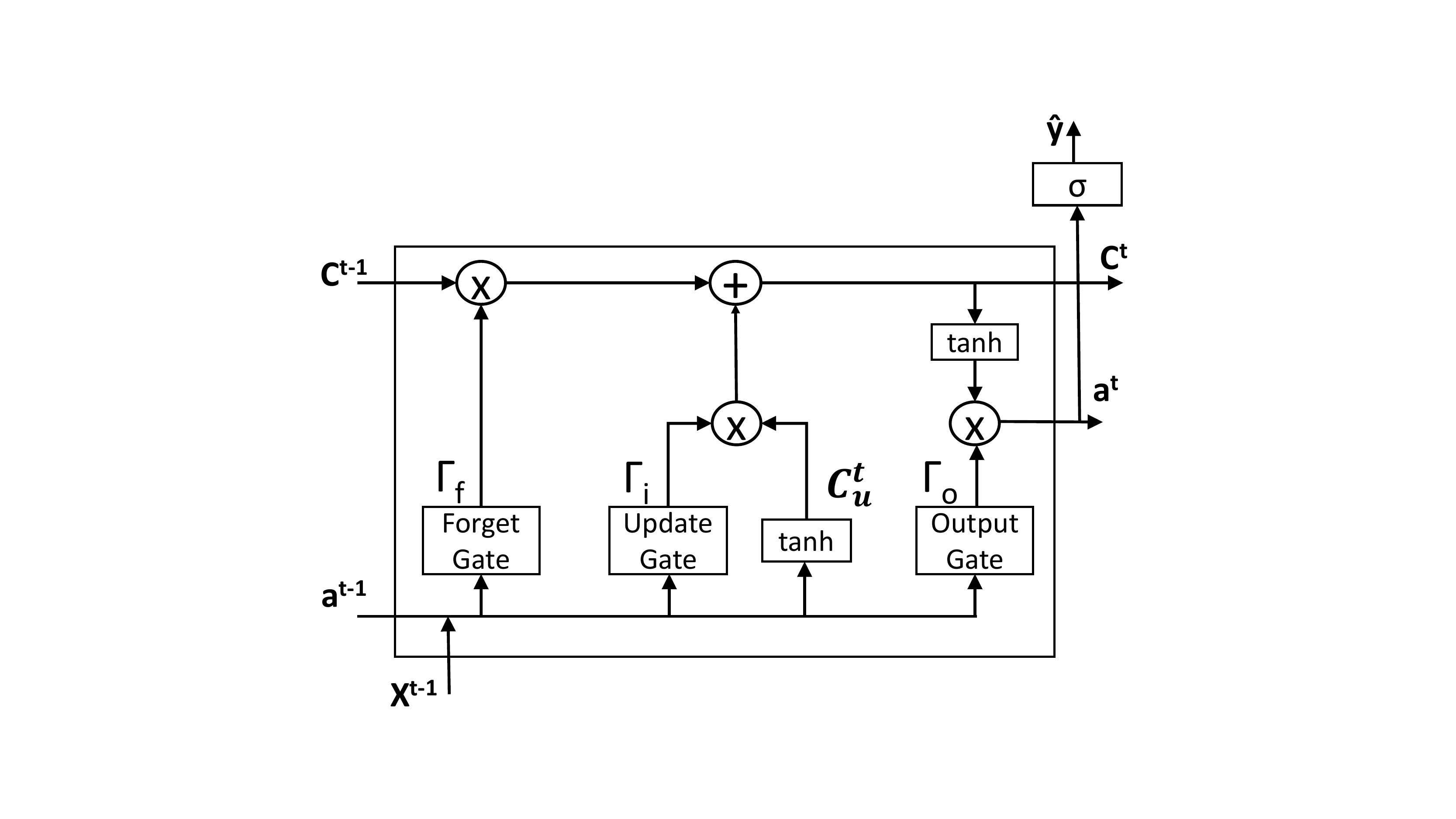}	
	\caption{Single LSTM Cell.}
	\label{fig:LSTM}
\end{figure}

The CDR does not specify activity in terms of particular units. However, an intuitive interpretations is that the activities are proportional to the amount of real traffic. For example, the magnitude of Inbound or outbound SMS activities are high for a greater number of SMS received or sent, respectively. The data was provided in the raw format. Hence, we will discuss the data cleansing method in the next step. 

\subsection{Data cleansing}
The CDR, in its raw format, could not be used to extract any meaningful information. Hence we applied data cleansing and filtering over the CDR. The timestamps were changed from milliseconds to minutes. There were some missing fields which we marked as zeros (0). There were multiple entry records for each timestamp. We summed them to make a single activity record per timestamp. 
Figure \ref{fig:Activty} shows the 24-hour Internet Activity for Grid 01. 

In our prediction model, we have only used Internet traffic Activity. However, the our model can  be used to predict activities for SMS and calls without any modification. We will discuss traffic prediction in the next section. 

\section{Cellular Traffic Prediction}

In this section, we will first briefly describe basics of feed forward and recurrent neural network (NN) followed by the LSTM based learning model. 

\label{sec_traffic_prediction}
\subsection{Feed Forward and Recurrent NN}

In artificial neural networks, the  nodes are connected to form a directed graph which is ideal for handling temporal sequence predictions. In vanilla feed forward network (FFNN), information flows only in forward direction. In FFNN, the input layer feeds the forward looking hidden layer for calculations and manipulations. The hidden layers forward the information to the output layer which produce regression of classification predictions.

\begin{figure}[t]
	\centering
	\includegraphics[width=3.8 in, trim={4.5cm 6.0cm 3cm 5.0cm},clip = true]{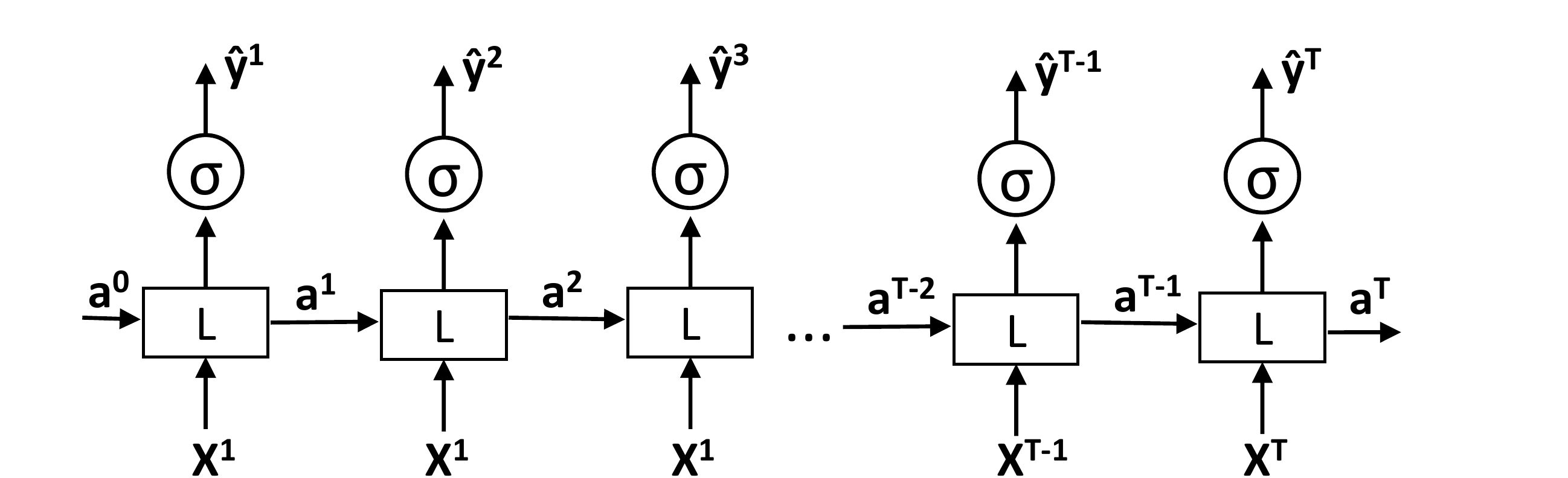}	
	\caption{LSTM Network.}
	\label{fig:LSTM_network}
\end{figure}

A NN maps inputs to the outputs by learning from the examples provided during the training phase and can be used for prediction in classification and regression problems. 
During the training process, the predictions are compared against the expected output values (often known as ground truth data) to calculate the loss function. In the beginning of the training, the loss function is usually quite high indicating the incorrect prediction by the model. With back propagation and gradient descent method, the model adjust the weights and biases corresponding to the input value to minimize the loss function. A fully trained NN has the minimal loss (also called as error) between the predicted and the expected output value \cite{goodfellow2016deep}. After successful training, the model is validated and a validation error is calculated. A model is fully trained for prediction when the training and validation errors are both minimized.

In recurrent neural network (RNN), though the learning process is the same as FFNN, the architecture is slightly different. RNNs takes the output of one layer, and feed it as the input to the next layer. Hence, each layer has information from the past input values. RNN considers the current input as well as the input received in the previous time steps during training and prediction. This enables RNN to learn the knowledge from the all previous time instances to make a well informed prediction for time series data. 


However, vanilla RNNs have inherent vanishing and exploding gradient problem which halts the learning process as gradient either diminishes completely or explodes to very large value. Hence Long-Short-Term-Memory (LSTM), which is a variant of RNN was proposed in \cite{hochreiter1997long}. LSTMs were designed to avoid the long-term dependency issue, which is the cause of the vanishing-gradient problem in Vanilla RNNs \cite{goodfellow2016deep}.

\subsection{Learning Through LSTMs}

The structure of LSTM units (often known as cells) enable a neural network to learn long term dependencies. The learning processing is strictly controlled by multiple gates that allow (or bar) the flow of incoming data from the previous cell and/or input as shown in Figure \ref{fig:LSTM}. The standard LSTM unit is shown in Figure \ref{fig:LSTM}. There are three main gates in any LSTM unit, the forget gate ($\Gamma_f$), the update or input gate ($\Gamma_i$), and the output gate ($\Gamma_o$). The cell state for the current unit $C^t$ is updated by the information passed through the update gate ($\Gamma_i$). The candidate value for current cell's state (i.e. $C_u^t$) is updated based on the information from the previous hidden state (i.e. $\text{a}^{t-1}$) and input $X^t$. The update gate decides to allow or bar the flow of this candidate value to the output state. Finally the output gate $\Gamma_o$ allows the information to pass from the current cell. The forget gate lets the current cell keep or forget the state value from the previous time step. The prediction is made as $\hat{y}$ after passing through an activation function (often sigmoid or softmax). 

\begin{figure}[t]
	\centering
	\includegraphics[width=4.0 in, trim={9.5cm 4.8cm 6.5cm 5.0cm},clip = true]{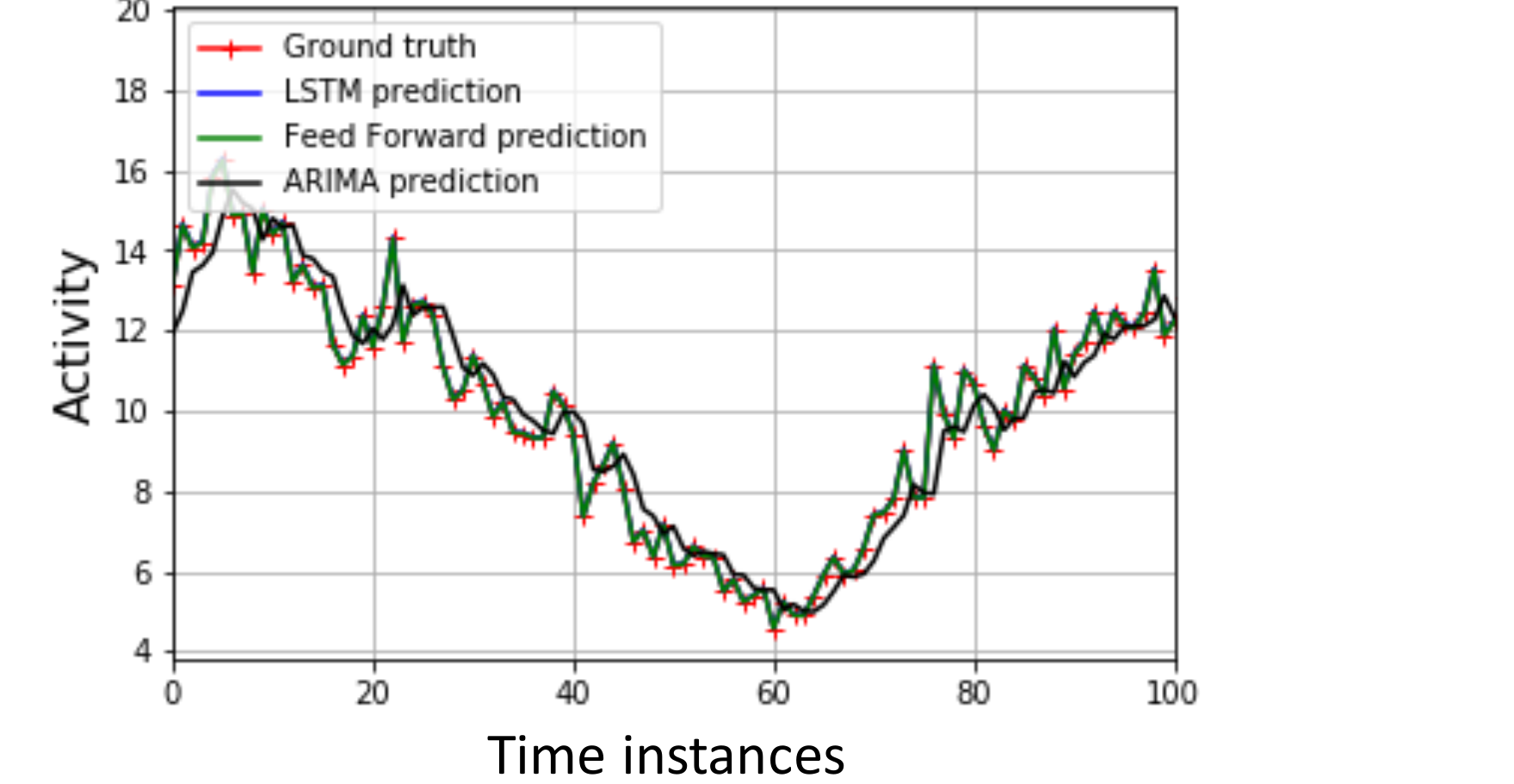}	
	\caption{Traffic prediction with large-sized training set.}
	\label{fig:combined_predict_80percent}
\end{figure}

The LSTM cells are chained to form one layer of the LSTM network as shown in Figure \ref{fig:LSTM_network}. Each cells computes operation for one time step and transfer the output to the next cell. The number of cells in a LSTM network indicates the amount of observations of data that is being considered before making any prediction.  For our case, the input $X^t$ is the internet activity and the number of observations is the amount of selected time steps T.     

The expression for all the gates, cell states, out of the hidden layer, and the final prediction are given as below:

\begin{equation}
\Large \Gamma_f^{t} = \sigma(W_f[a^{t-1}, X^t] + b_f)
\end{equation}

\begin{equation}
\Large \Gamma_i^{t} = \sigma(W_i[a^{t-1}, X^t] + b_i)
\end{equation}

\begin{equation}
\Large C_u^t = \varphi(W_c[a^{t-1}, X^t] + b_c)
\end{equation}

\begin{equation}
\Large \Gamma_o^{t} = \sigma(W_o[a^{t-1}, X^t] + b_o)
\end{equation}

\begin{equation}
\Large C^t =  \Gamma_f^t \ast C^{t-1} + \Gamma \ast C_u^t
\end{equation}

\begin{equation}
\Large a^t =  \Gamma_o^t\ast\varphi(C^t)
\end{equation}

The final output $Y^t$ is then calculated as :

\begin{equation}
\Large y^t = \sigma(W_y a^t + b_y) 
\end{equation}

In the equations above,  symbol $\sigma$ represent the sigmoid function which is often known as the squashing function because it limits the output between 0 (gate OFF) and 1 (Gate fully ON). Formally, the sigmoid function is defined as  \Large$ \sigma(x) = \Large \frac{1}{1 + e^{-x}}$ \normalsize. Symbol $\varphi$ is another squashing function and often $\tanh$ or rectified linear unit (relu) operations are used for $\varphi$. Readers can refer to relevant literature to gather further information about these functions. \cite{goodfellow2016deep}. The symbol $\ast$ represents the element-wise multiplication. Finally, $W_{(.)}$ and $b_{(.)}$ are the vectors of weights and biases corresponding to the respective gates, hidden layer, input, and output layer. The exact values of these weights and biases are selected by the libraries described in the next sub-section.

\subsection{Training and Prediction Software}
We have used Matlab for data cleansing and filtering. All the algorithms are implemented in Python using Keras and Scikit libraries with Tensorflow at the backend. 

\begin{figure}[t]
	\centering
	\includegraphics[width= \linewidth, trim={0cm 0cm 0cm .65cm},clip = true]{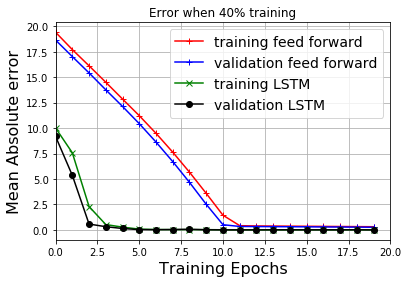}	
	\caption{Error for traffic prediction with medium-sized training set.}
	\label{fig:combined_error_40percent}
\end{figure}

\section{Results and Discussion}
\label{sec_results}
In this section we will show the performance comparison of LSTM model with the base line ARIMA and vanilla feed forward neural network model. 
We have compared the performance of each technique with the ground truth test data from CDR. We have fixed the training epochs to 20 for each cycle. 
For LSTM model, we have used two hidden layers to make an even comparison with the FFNN and ARIMA. The first hidden layer contains 50 LSTM cells followed by a dense layer with single unit. The FFNN contains two hidden layers with first layer containing 5 activation units activated by relu operation. The second hidden layer contains one non-linear activation unit. Training and validation losses are calculated using mean absolute error. 
\begin{figure}[t]
	\centering
	\includegraphics[width=4.1 in, trim={9.5cm 4.8cm 6.5cm 5.0cm},clip = true]{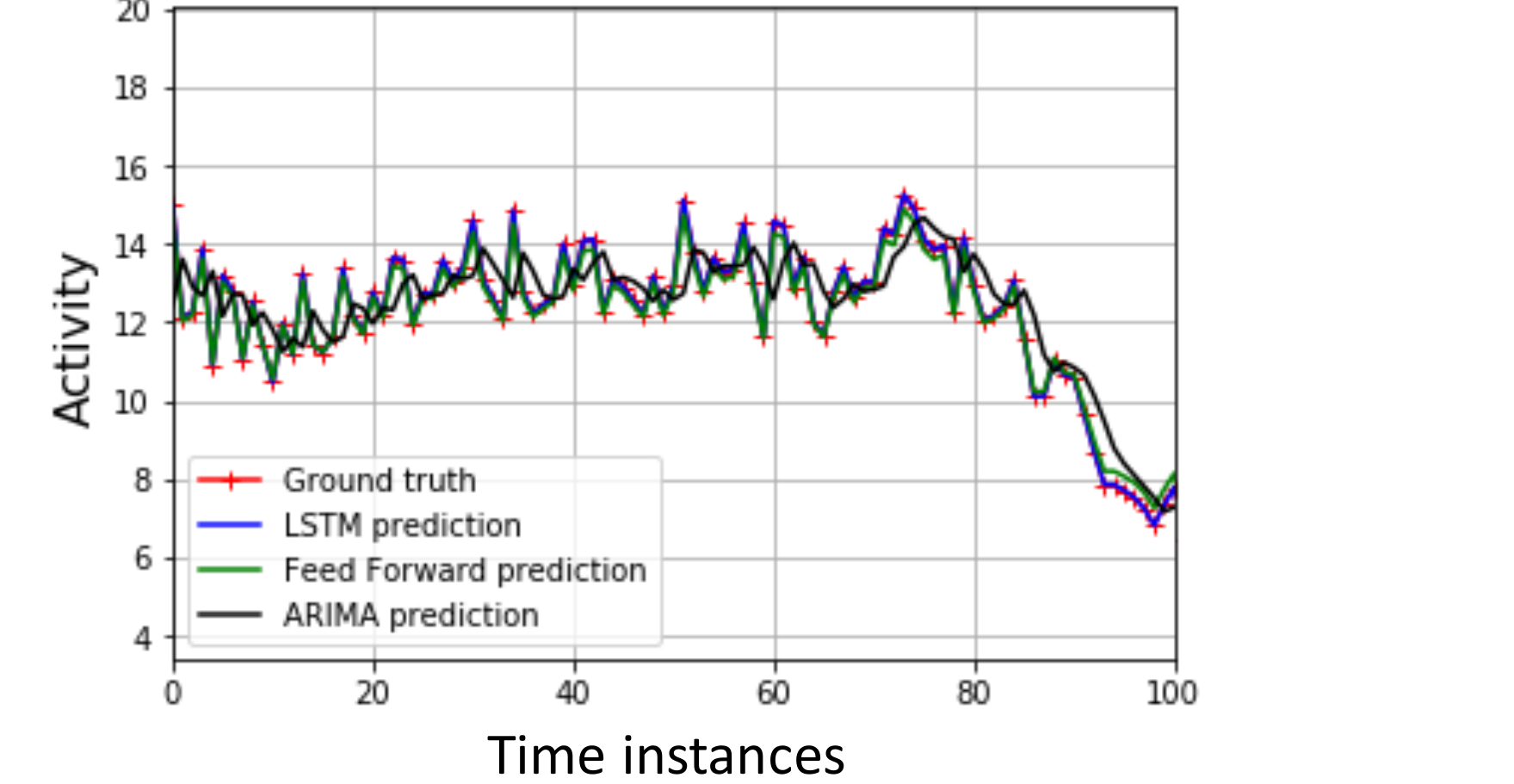}	
	\caption{Traffic prediction with medium-sized training set.}
	\label{fig:combined_predic_40percent}
\end{figure}

Figure \ref{fig:combined_predict_80percent} shows the traffic prediction by LSTM, FFNN model, and ARIMA model. We used 7142 samples for training LSTM and FFNN models. For validation and testing, we used 893 samples for each case. 
The LSTM and FFNN both learned the pattern in less than 5 epochs due to large number of training examples. 
It can be observed that LSTM and FFNN predictions match to that of the ground truth data. 
The ARIMA model predicts very close to the ground truth but does not exactly match the traffic pattern.  
\begin{figure}[!t]
	\centering
	\includegraphics[width=4.1 in, trim={9.5cm 4.8cm 6.5cm 5.0cm},clip = true]{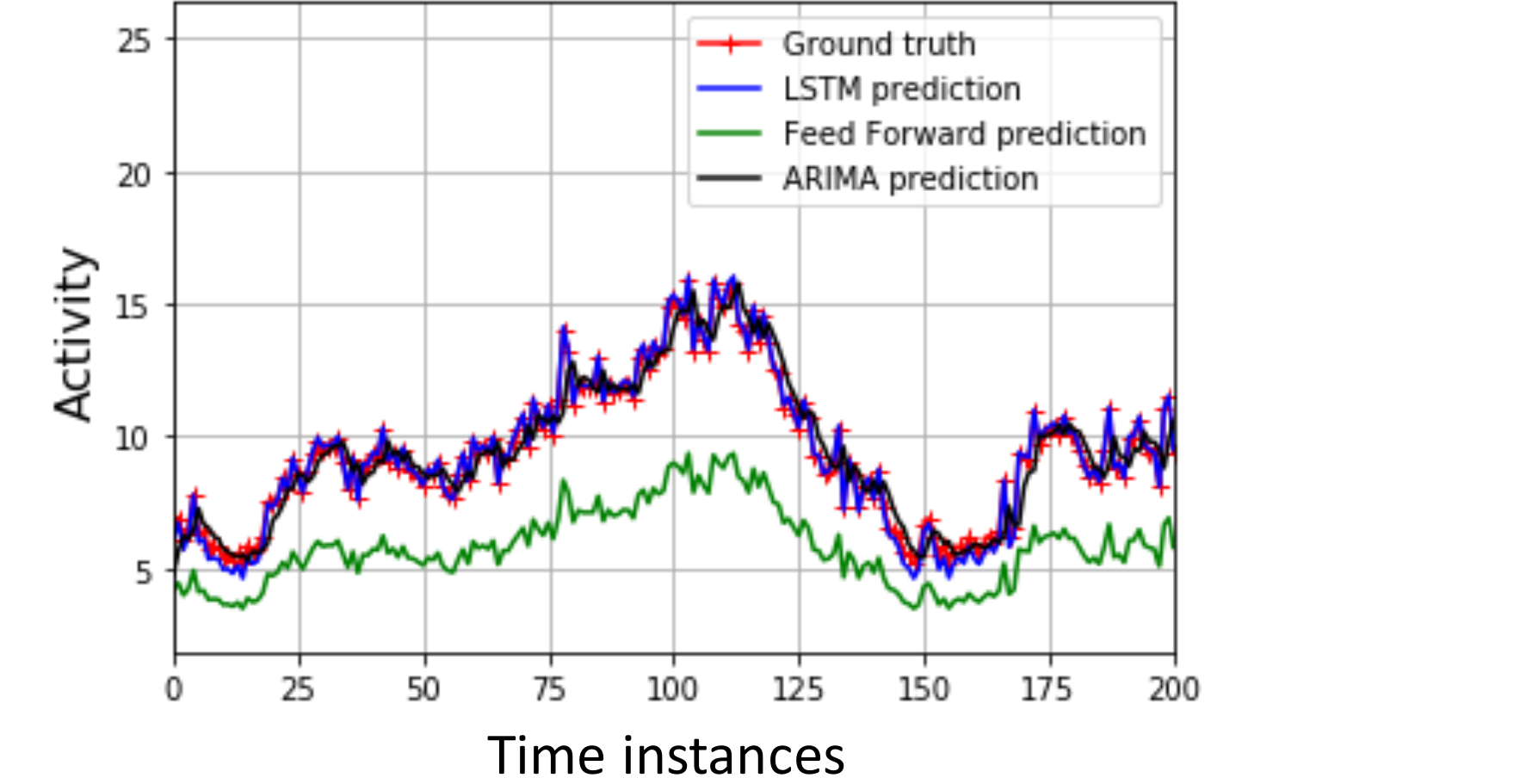}	
	\caption{Traffic prediction with small-sized training set.}
	\label{fig:combined_predict_10percent}
\end{figure}

We later reduced the training samples to 3571. The training and validation errors for this case are shown in Figure \ref{fig:combined_error_40percent} and the prediction results are presented in Figure \ref{fig:combined_predic_40percent}. We can observe that LSTM and FFNN still predict very accurately. The ARIMA baseline model however does not exactly match the ground truth traffic. It should be noted from Figure \ref{fig:combined_error_40percent} that when we reduced the number of training samples, the training and validation error for LSTM converges to near zero (0) only after 2 epochs. However, FFNN took at least 10 epochs to fully train the model to enable accurate predictions. Nevertheless, both models' errors converged to zero before the 20 epochs limit. 

\begin{figure}[!t]
	\centering
	\includegraphics[width= \linewidth, trim={0.0cm 0.0cm 0.0cm 0.7cm},clip = true]{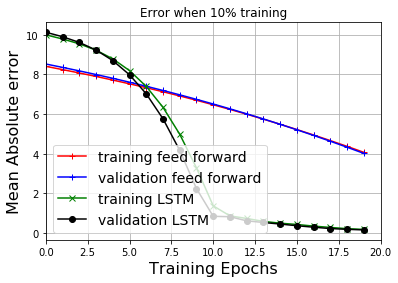}	
	\caption{Error for traffic prediction with small-sized training set.}
	\label{fig:combined_error_10percent}
\end{figure}

When we further reduced the training samples to 892, we observed that after training the models for 20 epochs, the FFNN could not predict according to actual ground truth data. In fact, its performance worsened even to that of the ARIMA model. The LSTM, on the other hand, very accurately predicted the traffic activity. This is due to the fact that LSTM trained the network within 20 epochs and the training and validation error converged to zero as shown in Figure \ref{fig:combined_error_10percent}. On the other hand, the error for the FFNN remains high even after 20 epochs. Interestingly, FFNN could estimate patterns of future traffic, however, with very low accuracy.

\section{Conclusion} 
\label{sec_conclusion}
In this paper, we presented cellular data traffic prediction using recurrent neural network, in particular, with long-short-term-memory model. We demonstrated that LSTM and vanilla feed-forward  neural networks predict more accurately as compared to the statistical ARIMA model. However, the  LSTM models were shown to be learning more quickly as compared to the FFNN, even with a small amount of training data sample. As our future work, we are working to design a LSTM based resource allocation method for 6G networks. 

\bibliographystyle{./bibliography/IEEEtran}
\bibliography{test_bib}

\begin{thebibliography}{10}
\providecommand{\url}[1]{#1}
\csname url@samestyle\endcsname
\providecommand{\newblock}{\relax}
\providecommand{\bibinfo}[2]{#2}
\providecommand{\BIBentrySTDinterwordspacing}{\spaceskip=0pt\relax}
\providecommand{\BIBentryALTinterwordstretchfactor}{4}
\providecommand{\BIBentryALTinterwordspacing}{\spaceskip=\fontdimen2\font plus
\BIBentryALTinterwordstretchfactor\fontdimen3\font minus
  \fontdimen4\font\relax}
\providecommand{\BIBforeignlanguage}[2]{{%
\expandafter\ifx\csname l@#1\endcsname\relax
\typeout{** WARNING: IEEEtran.bst: No hyphenation pattern has been}%
\typeout{** loaded for the language `#1'. Using the pattern for}%
\typeout{** the default language instead.}%
\else
\language=\csname l@#1\endcsname
\fi
#2}}
\providecommand{\BIBdecl}{\relax}
\BIBdecl

\bibitem{cerwall2018ericsson}
P.~Cerwall, A.~Lundvall, P.~Jonsson, R.~M{\"o}ller, S.~B{\"a}vertoft,
  S.~Carson, and I.~Godor, ``Ericsson mobility report 2018,'' 2018.

\bibitem{jaffry2018effective}
S.~Jaffry, S.~F. Hasan, and X.~Gui, ``Effective resource sharing in mobile-cell
  environments,'' \emph{arXiv preprint arXiv:1808.01700}, 2018.

\bibitem{jaffry2018shared}
------, ``Shared spectrum for mobile-cell's backhaul and access link,'' in
  \emph{2018 IEEE Global Communications Conference (GLOBECOM)}.\hskip 1em plus
  0.5em minus 0.4em\relax IEEE, 2018, pp. 1--6.

\bibitem{jaffry2017distributed}
S.~Jaffry, S.~F. Hasan, X.~Gui, and Y.~W. Kuo, ``Distributed device discovery
  in prose environments,'' in \emph{TENCON 2017-2017 IEEE Region 10
  Conference}.\hskip 1em plus 0.5em minus 0.4em\relax IEEE, 2017, pp. 614--618.

\bibitem{lecun2015deep}
Y.~LeCun, Y.~Bengio, and G.~Hinton, ``Deep learning,'' \emph{nature}, vol. 521,
  no. 7553, pp. 436--444, 2015.

\bibitem{voulodimos2018deep}
A.~Voulodimos, N.~Doulamis, A.~Doulamis, and E.~Protopapadakis, ``Deep learning
  for computer vision: A brief review,'' \emph{Computational intelligence and
  neuroscience}, vol. 2018, 2018.

\bibitem{ravi2016deep}
D.~Rav{\`\i}, C.~Wong, F.~Deligianni, M.~Berthelot, J.~Andreu-Perez, B.~Lo, and
  G.-Z. Yang, ``Deep learning for health informatics,'' \emph{IEEE journal of
  biomedical and health informatics}, vol.~21, no.~1, pp. 4--21, 2016.

\bibitem{deng2013new}
L.~Deng, G.~Hinton, and B.~Kingsbury, ``New types of deep neural network
  learning for speech recognition and related applications: An overview,'' in
  \emph{2013 IEEE International Conference on Acoustics, Speech and Signal
  Processing}.\hskip 1em plus 0.5em minus 0.4em\relax IEEE, 2013, pp.
  8599--8603.

\bibitem{young2018recent}
T.~Young, D.~Hazarika, S.~Poria, and E.~Cambria, ``Recent trends in deep
  learning based natural language processing,'' \emph{ieee Computational
  intelligenCe magazine}, vol.~13, no.~3, pp. 55--75, 2018.

\bibitem{zappone2019wireless}
A.~Zappone, M.~Di~Renzo, and M.~Debbah, ``Wireless networks design in the era
  of deep learning: Model-based, ai-based, or both?'' \emph{arXiv preprint
  arXiv:1902.02647}, 2019.

\bibitem{chen2019artificial}
M.~Chen, U.~Challita, W.~Saad, C.~Yin, and M.~Debbah, ``Artificial neural
  networks-based machine learning for wireless networks: A tutorial,''
  \emph{IEEE Communications Surveys \& Tutorials}, vol.~21, no.~4, pp.
  3039--3071, 2019.

\bibitem{shu2005wireless}
Y.~Shu, M.~Yu, O.~Yang, J.~Liu, and H.~Feng, ``Wireless traffic modeling and
  prediction using seasonal arima models,'' \emph{IEICE transactions on
  communications}, vol.~88, no.~10, pp. 3992--3999, 2005.

\bibitem{qiu2018spatio}
C.~Qiu, Y.~Zhang, Z.~Feng, P.~Zhang, and S.~Cui, ``Spatio-temporal wireless
  traffic prediction with recurrent neural network,'' \emph{IEEE Wireless
  Communications Letters}, vol.~7, no.~4, pp. 554--557, 2018.

\bibitem{zhao2019celltrademap}
Y.~Zhao, Z.~Zhou, X.~Wang, T.~Liu, Y.~Liu, and Z.~Yang, ``Celltrademap:
  Delineating trade areas for urban commercial districts with cellular
  networks,'' in \emph{IEEE INFOCOM 2019-IEEE Conference on Computer
  Communications}.\hskip 1em plus 0.5em minus 0.4em\relax IEEE, 2019, pp.
  937--945.

\bibitem{azari2019cellular}
A.~Azari, P.~Papapetrou, S.~Denic, and G.~Peters, ``Cellular traffic prediction
  and classification: a comparative evaluation of lstm and arima,'' in
  \emph{International Conference on Discovery Science}.\hskip 1em plus 0.5em
  minus 0.4em\relax Springer, 2019, pp. 129--144.

\bibitem{elnashar2014design}
A.~ElNashar, M.~A. El-Saidny, and M.~Sherif, \emph{{Design, deployment and
  performance of 4G-LTE networks: A practical approach}}.\hskip 1em plus 0.5em
  minus 0.4em\relax John Wiley \& Sons, 2014.

\bibitem{telecom_ItaliaCDR}
\BIBentryALTinterwordspacing
``{Telecom Italia, Open Big Data, Milano Grid},'' Online, 2014. [Online].
  Available: \url{https://dandelion.eu/}
\BIBentrySTDinterwordspacing

\bibitem{parwez2017big}
M.~S. Parwez, D.~B. Rawat, and M.~Garuba, ``Big data analytics for
  user-activity analysis and user-anomaly detection in mobile wireless
  network,'' \emph{IEEE Transactions on Industrial Informatics}, vol.~13,
  no.~4, pp. 2058--2065, 2017.

\bibitem{goodfellow2016deep}
I.~Goodfellow, Y.~Bengio, and A.~Courville, \emph{Deep learning}.\hskip 1em
  plus 0.5em minus 0.4em\relax MIT press, 2016.

\bibitem{hochreiter1997long}
S.~Hochreiter and J.~Schmidhuber, ``Long short-term memory,'' \emph{Neural
  computation}, vol.~9, no.~8, pp. 1735--1780, 1997.

\end{thebibliography}

\end{document}